\documentclass[journal=jacsat,manuscript=article]{achemso}

\usepackage{chemformula} 
\usepackage[T1]{fontenc} 
\usepackage{graphicx}
\usepackage{dcolumn}
\usepackage{bm}
\usepackage{mathtools}
\usepackage{mathrsfs, amsmath, wasysym}
\usepackage{caption}
\usepackage[mathscr]{eucal}
\usepackage{xfrac}
\usepackage{graphicx}
\usepackage{dcolumn}
\usepackage{bm}
\usepackage{color}
\usepackage{tabularx}

\author{Zhenyao Fang}
\affiliation[Northeastern University]{Department of Physics, Northeastern University, Boston, MA 02115, USA}
\email{z.fang@northeastern.edu}
\author{Ting-Wei Hsu}
\affiliation[Northeastern University]{Department of Physics, Northeastern University, Boston, MA 02115, USA}
\author{Qimin Yan}
\affiliation[Northeastern University]{Department of Physics, Northeastern University, Boston, MA 02115, USA}
\email{q.yan@northeastern.edu}

\title{A Machine Learning Framework for Modeling Ensemble Properties of Atomically Disordered Materials}

\begin{document}

\begin{abstract}
Disorder, though naturally present in experimental samples and strongly influencing a wide range of material phenomena, remains underexplored in first-principles studies due to the computational cost of sampling the large supercell and configurational space. The recent development of machine learning techniques, particularly graph neural networks (GNNs), has enabled the efficient and accurate predictions of complex material properties, offering promising tools for studying disordered systems. In this work, we introduce a computational framework that integrates GNNs with Monte Carlo simulations for efficient calculations of thermodynamic properties and ensemble-averaged functional properties of disordered materials. Using the surface-termination-disordered MXene monolayer \ch{Ti3C2T}$_{2-x}$ as a representative system, we investigate the effect of surface termination disorder involving \ch{-F}, \ch{-O}, and termination vacancies on the electrical and optical conductivity spectra. We find that surface termination disorder affects the temperature dependence of electrical conductivity, inducing a peak close to the order-disorder phase transition temperature that reflects the competition between scattering and electron filling effects of the surface termination groups across the phase transition. In contrast, optical conductivity remains robust to local disorder across a wide temperature range and is governed primarily by the global chemical composition of surface terminations. These results demonstrate the utility of our machine-learning-assisted framework for statistically modeling disorder effects and ensemble properties in complex materials, opening new avenues for future studies of disorder-driven phenomena in systems such as high-entropy alloys and disordered magnetic compounds.
\end{abstract}

\section*{Introduction}
Disorder plays a crucial role in determining the physical and chemical behavior of materials, including localized electronic states~\cite{Pu22DisorderImpactElectronic, Cutler69DisorderImpactElectronic, Anderson58DisorderImpactElectronic}, modulated lattice dynamics~\cite{Alam21DisorderImpactLattice, Wang20DisorderImpactLattice, Huang21DisorderImpactLattice}, and enhanced catalytic activity~\cite{Xie21DisorderImpactCatalytic, Sharma21DisorderImpactCatalytic}. Based on whether the chemical bonding network is disrupted or not, disorder in materials is typically categorized into configurational disorder and structural disorder. Configurational disorder is characterized by each atomic site being occupied by different chemical species (compositional disorder or chemical disorder) or spin degrees of freedom (spin disorder)~\cite{Cordell21ConfigurationalDisorder}, while structural disorder includes disruptions to the bonding network such as vacancy defects and planar dislocations~\cite{Cliffe10StructuralDisorder, Leem24WS2}. These local atomic variations can collectively modulate macroscopic material properties that are directly measured in experiments, such as mechanical strength in high-entropy alloys, electrical and thermal transport in thermoelectric materials, and catalytic activity in heterogeneous catalysts.

Despite its significance, accurate modeling the effect of disorder on material properties still remains challenging. Both configurational and structural disorder require sampling over large supercells and vast configuration spaces to obtain the thermodynamic average of macroscopic properties~\cite{Prokhorenko18MetropolisSampling, Wang01WLSampling, Zhou06WLSampling}. Such ensemble averaging is essential for meaningful comparison with experiments, which are typically performed at finite temperatures and inherently reflect ensemble-averaged properties instead of properties of a single idealized configuration. This makes direct first-principles approaches computationally prohibitive. To address those limitations, computational approaches such as cluster expansion methods~\cite{Seko09CE, Drautz19CE, Sanchez93CE} and large-scale molecular dynamic simulations~\cite{Choyal24MLIP, Anstine23MLIP, Casillas24MLIP} have been developed to accelerate the sampling process. However, they were primarily applied to properties closely related to crystal structures, such as electric polarization, phase transition, and short-range ordering. In contrast, they are less effective for properties such as optical and transport responses, which are primarily governed by the electronic structure and cannot be inferred directly from the crystal structure.

Recent years, machine learning techniques, particularly graph neural networks (GNNs), have emerged as powerful tools for predicting material properties~\cite{Xie18CGCNN, Petar18GAT, Fung21GNN, Reiser22GNN}. By representing atomic structures as graphs, GNNs can effectively learn the chemical environment of each atom and aggregate those information to predict a wide range of target properties, including scalar quantities (such as formation energy, band gap, and defect formation energy)~\cite{Xie18CGCNN, Fang25Persistent}, sequential quantities (such as the electronic and phonon density of states, and the absorption coefficient spectrum)~\cite{Bai22xtaldos, Chen21PhononDOS, Hung24GNNOpt}, and tensorial properties (such as the elastic properties and the Hamiltonian)~\cite{Hestroffer23GNNMechanical, LiHe2022DeepH}. In previous studies, GNNs have also shown promise in modeling the energetics and thermodynamic properties of configurationally disordered alloys~\cite{Fang24DisorderGNN}. However, their full potential to capture the combined effects of configurational and structure disorder on the ensemble average of functional properties, particularly those closely related to the electronic structure, remains largely unexplored. 

In this work, we introduce a statistical computational framework that integrates GNNs and Monte Carlo (MC) simulations to model atomically disordered materials and compute the thermodynamic averages of functional properties that are inaccessible to first-principles methods. The GNN is trained to predict the energy and other physical quantities, such as the density of states or the conductivity spectrum, from atomic configurations, and is then embedded within MC simulations to enable efficient sampling of configurations at finite temperatures. As a case study, we focus on the MXene monolayer \ch{Ti3C2T}$_x$, where atomic disorder arises from the mixture of surface terminations \ch{T}$_x$, most commonly \ch{-O}, \ch{-F}, and termination vacancies~\cite{Fang24MXene, Lim22MXene, Vahid21MXene, Hu17MXeneSurfaceTerminationOrigin}. These surface terminations significantly affect the electronic structure and further electrical transport and optical responses, making MXenes an ideal platform to demonstrate the capability of the proposed framework. Our results reveal that the order-disorder phase transition temperature increases with the atomic fraction of \ch{-F} groups and termination vacancies. More importantly, we find that electrical conductivity exhibits a peak near the phase transition temperature, which is absent for pristine MXenes. This peak arises from the competition between the scattering effects and electron filling effects from the \ch{-F} groups, suggesting distinct charge transport mechanism in the ordered and disordered phase. On the contrary, the optical conductivity remains largely unaffected by surface termination disorder and is mainly governed by the chemical composition of the surface termination groups. In particular, we observe a clear correlation between the intensity of the optical conductivity peak at 1.5~eV and the fraction of \ch{-F} groups. These findings provide new insights into how surface termination disorder can affect the physical responses of MXenes, and highlight our framework as a data-efficient and universal approach to connect local atomic orderings to global macroscopic measurable quantities, offering a new paradigm for disorder-aware materials modeling.

\section*{Results and Discussions}
\paragraph{Statistical Modeling of Ensemble Properties}
The accurate modeling of the thermodynamic and ensemble properties of disordered materials requires a statistical treatment of the large configurational space. In principle, ensemble properties at finite temperatures can be calculated using methods such as MC simulations. However, first-principles methods, such as density-functional theory (DFT), are computationally too expensive to sample a large number of configurations and to compute complex properties such as conductivity spectra, motivating the development for machine-learning-assisted approaches that can make efficient and accurate predictions.

\begin{figure}[htb]
\centering
\includegraphics[width=\linewidth]{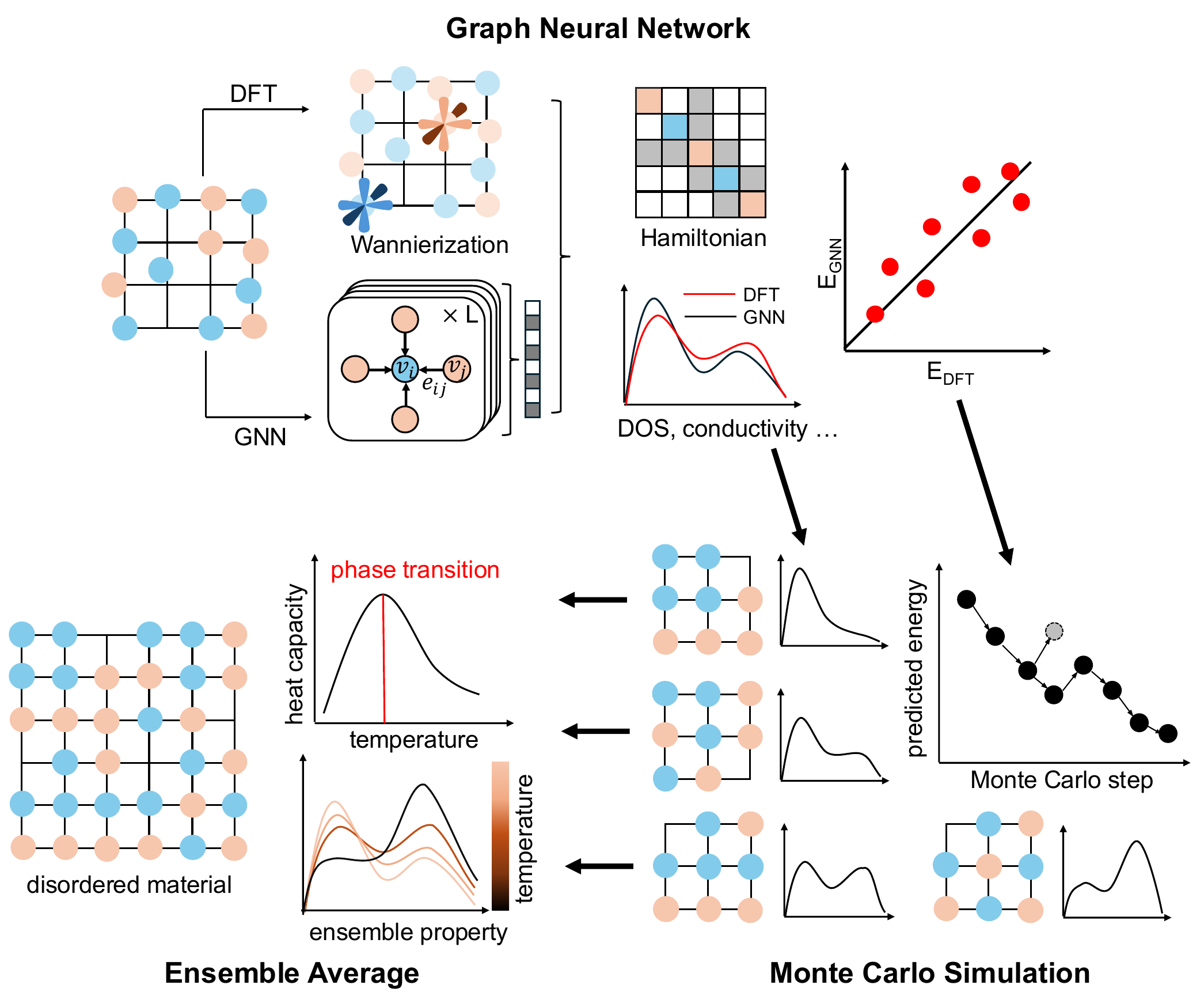}
\caption{A schematic plot of the computational framework to integrate GNN and MC simulations to compute the ensemble properties of disordered materials. The GNN model is trained a dataset of configurations, whose functional properties are calculated through a high-throughput workflow based on maximally localized Wannier functions. The GNN model is applied to MC simulation for efficient sampling of the configurational space, and the output is used to compute the thermodynamic properties, such as the order-disorder phase transition temperature and the ensemble average of functional properties, such as the conductivity spectrum.}
\label{fig:workflow}
\end{figure}

To enable such framework, a dataset must first be constructed to represent the relevant configurational space, as shown in Figure~\ref{fig:workflow}. In this work, we employed a high-throughput workflow~\cite{Fang25Database} based on maximally localized Wannier functions~\cite{Marzari12WannierFunctions, Zhang18HighThroughputWannierization} to calculate complex physical properties such as the conductivity spectra. The workflow automates the analyze the projected density of states and selects the chemical orbitals with leading contributions to the states near the Fermi level as the projection functions. It also determines the energy windows by an energy scan based on the density of states analysis. As demonstrated in our previous work~\cite{Fang25Database}, this workflow can produce well-localized Wannier functions and reproduce the DFT band structures near the Fermi level with errors typically within a few meV for a wide range of materials. The resulting Wannier tight-binding Hamiltonian allows for efficient evaluations of functional properties, such as density of states, optical responses, and transport properties, through Wannier interpolation scheme~\cite{Marzari12WannierFunctions}.

GNNs are then trained on this dataset to learn the structure-property relation for each configuration, as shown in Figure~\ref{fig:workflow}. In these models, the crystal structure is represented as a graph, with atoms as nodes and chemical bonds as edges. The node features typically consist of the one-hot encoding of the atomic species~\cite{Xie18CGCNN}, while the edge features are constructed from the interatomic distances expanded in a Gaussian basis. The graph passes through several stacked graph convolution layers that update node features by aggregating information from nearby nodes weighted by the edge features; for example, attention-based GNNs enable flexible message-passing mechanism that can adaptively adjust neighboring contributions~\cite{GAT1, GAT2, Shi21Transformer}, while equivariant GNNs encode the physical symmetries (rotational and translational symmetries) into the model, making them particularly suited for material property predictions~\cite{Batzner22EquivariantGNN, Hung24GNNOpt}. After those convolution operations, the node features are gathered into a graph-level global feature, which is mapped to the target properties through multi-layer perceptrons. 

Once trained, the GNN is applied to evaluate energies and other target properties for each sampled configuration during the MC simulation. The choice of ensemble depends on whether atomic species can exchange with a reservoir, and different sampling methods could be used to explore the configurational space. The Wang-Landau algorithm, for example, can efficiently calculate the configurational density of states, which can be used to calculate thermodynamic properties at arbitrary temperatures~\cite{Wang01WLSampling, Fang24DisorderGNN}. However, in this work, we employed the Metropolis sampling because it allows straightforward evaluation of the thermodynamic average of both energies and and other functional properties. This framework thus enables the quantitative and data-efficient analysis of disorder effects in complex materials with high accuracy but significantly lower computational cost.

\paragraph{Fully Terminated MXene \ch{Ti3C2T2}}
Two-dimensional MXenes have attracted considerable attention due to their chemical versatility, tunable electronic structures, and wide applications in photovoltaics, topological materials, and catalysts~\cite{Shahzad16MXene, Lyu19MXene, Naguib11MXene, Gogotsi19MXene,  Zhao22MXeneApplication, Zhang22MXeneApplication}. Among them, \ch{Ti3C2T}$_{2-x}$ is widely studied as a prototypical system, whose functional properties have been extensive characterized in experiments~\cite{Fang24MXene, Fang24MXeneTransport}. Here, \ch{T}$_{2-x}$ represents the surface termination groups, typically including \ch{-O}, \ch{-F}, \ch{-OH}, \ch{-Cl}, \ch{-H} functional groups, and $x > 0$ reflects the presence of termination vacancies. Among these surface termination groups, \ch{-O} and \ch{-F} are the most thermodynamically stable ones and frequently coexist in experimentally synthesized samples~\cite{Caffrey18MXeneSurfaceTermination, Dahlqvist24MXeneSurfaceTermination, Hu17MXeneSurfaceTerminationOrigin}.

\begin{figure}[htb]
\centering
\includegraphics[width=\linewidth]{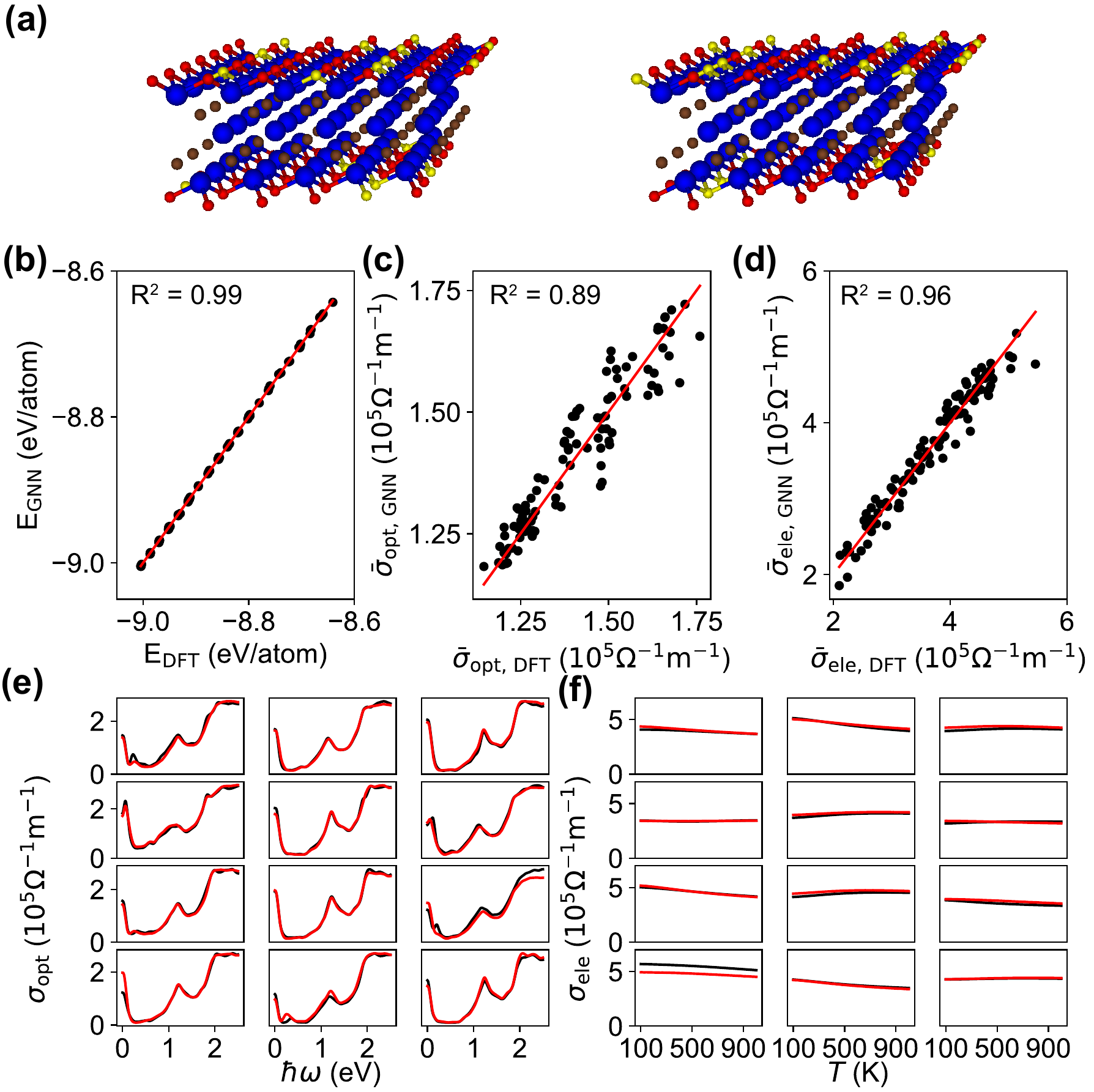}
\caption{(a) Examples of configurations in the fully terminated MXene \ch{Ti3C2O}$_{2-x}$\ch{F}$_{x}$ dataset. Color code: blue (Ti), brown (C), red (O), yellow (F). (b, c, d) The comparison of the energy (b), reduced optical conductivity (c), and reduced electrical conductivity (d) between the DFT and GNN methods. (e, f) The comparison of the optical conductivity (e) and electrical conductivity (f) spectra between the DFT (black) and GNN (red) methods for configurations in the test set.}
\label{fig:fully_terminated_GNN}
\end{figure}

To study the effect of surface termination disorder, we first considered the simplified case of fully terminated \ch{Ti3C2T2} structures, excluding termination vacancies. As described in the framework above, we constructed a dataset consisting 1,000 \ch{Ti3C2O}$_{2-x}$\ch{F}$_{x}$ configurations within a $5 \times 5 \times 1$ supercell with 175 atoms, where the amount of \ch{-F} group was randomly sampled in the range of $0 < x \leq 1$. Examples of the configurations can be found in Figure.~\ref{fig:fully_terminated_GNN}(a). Through a high-throughput workflow based on maximally localized Wannier functions~\cite{Fang25Database}, we calculated the energy $E$, the optical conductivity $\sigma_\text{opt} (\hbar \omega)$ (with the photon energy $\hbar \omega$ ranging from 0 to 2.5~eV) and the electrical conductivity $\sigma_\text{ele} (T)$ (with the temperature $T$ ranging from 100 to 1,000~K) for each configuration.
Furthermore, to ensure balanced optimization across these three targets, we normalize the energies to the range of $\left[0, 1\right]$ by min-max scaling, and normalize the optical and electrical conductivities by their respective mean values over the whole dataset (the details and the effect of other normalization schemes can be found in the Supplementary Information S2). In the following, however, all metrics are reported based on the unnormalized targets.

Based on the constructed dataset, we constructed and trained an equivariant GNN to simultaneously predict the energy, optical conductivity, and electrical conductivity of each atomic configuration~\cite{Hung24GNNOpt}. We chose the equivariant GNN architecture because of its sensitivity to variations in local atomic orderings and effectiveness in predicting sequential target properties~\cite{Batzner22EquivariantGNN, Hung24GNNOpt}. In Figure~\ref{fig:fully_terminated_GNN}, we present the comparison between the DFT ground truth and the predictions from the GNN model. For the spectral targets (optical and electrical conductivities), we reduce each spectrum to its mean value over their respective domain (photon energy for optical conductivity, and temperature for electrical conductivity), denoted as $\bar{\sigma}$ (see Supplementary Information S3 for details). The comparison between the DFT calculated and GNN predicted energies, reduced optical and electrical conductivities are shown in Figure.~\ref{fig:fully_terminated_GNN}(b-d), respectively. The coefficients of determination $(R^2)$ on the test set are 0.99, 0.89, 0.96 for energy, optical conductivity, and electrical conductivity, respectively. Besides, the corresponding mean absolute percentage errors are 0.1\%, 2.7\%, 3.8\%, indicating the high performance of our model to predict the three targets (the definitions of these metrics can be found in the Supplementary Information S3). Additionally, in Figure~\ref{fig:fully_terminated_GNN}(e, f), we randomly chose 12 configurations from the test set and show the comparison between the DFT calculated (black lines) and GNN predicted (red lines) optical and electrical conductivity spectra. The close agreement observed in both the energy and the conductivity spectra further supports the reliability of our GNN model for use in downstream MC simulations. Importantly, once trained, the GNN model predicts those properties of a given configuration within milliseconds, offering a significant speedup compared to DFT calculations. This computational efficiency, combined with its high accuracy, makes GNN models promising tools for predicting complex material properties and capturing disorder effects.

With the GNN model trained on the \ch{Ti3C2O}$_{2-x}$\ch{F}$_{x}$ dataset, we performed MC simulations to unveil the effect of surface termination disorder on the thermodynamic properties and ensemble average of electrical and optical conductivities. Using the Metropolis sampling algorithm within the canonical ensemble, we calculated the configurational heat capacity and the thermodynamic averages of electrical and optical conductivities, defined by Equations~\ref{eqn:heat_capacity} and~\ref{eqn:conductivity_average}. Note that while the electrical conductivity $\sigma_\text{ele} (T)$ is intrinsically temperature-dependent and in principle does not require the whole spectrum to compute its average at a specific temperature, we still chose to predict the full electrical conductivity spectrum in our MC simulations. This approach allows us to fully exploit the sequential and continuous nature of the spectrum as captured by the GNN model. Besides, for electrical and optical conductivity, we computed the thermodynamic average using the harmonic mean of the conductivities across sampled configurations. This choice is motivated by viewing the ensemble as a stack of configurational supercells, effectively resembling conductors connected in series, whose total conductivity is the harmonic mean of each individual conductor.

\begin{figure}[htb]
\centering
\includegraphics[width=\linewidth]{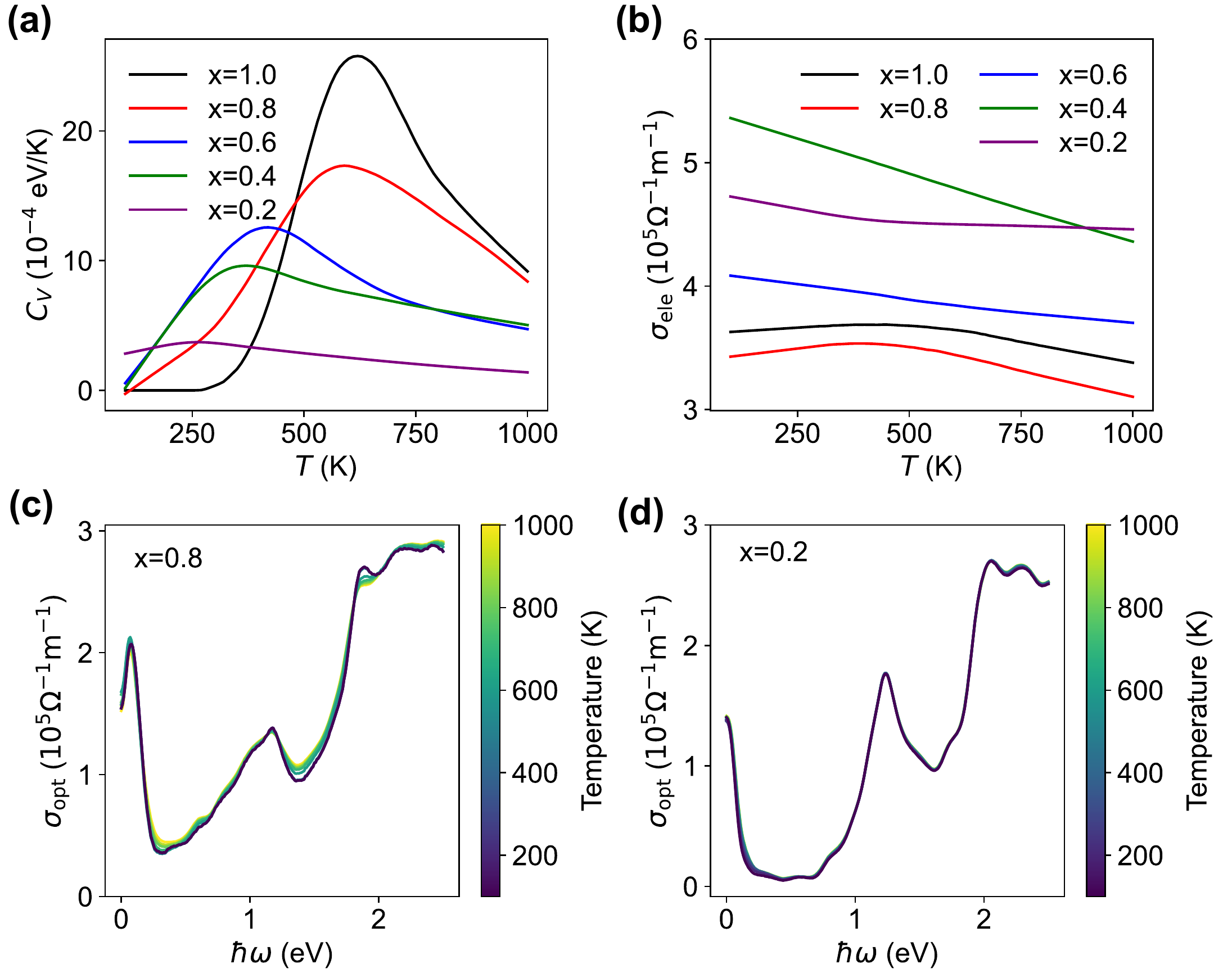}
\caption{The thermodynamic property and the ensemble averaged electrical and optical conductivities of the fully terminated MXene \ch{Ti3C2O}$_{2-x}$\ch{F}$_{x}$ under grand canonical ensemble. (a) The calculated configurational heat capacity at different \ch{-F} stoichiometries. (b) The thermodynamic average of the electrical conductivity at different stoichiometries. (c, d) The thermodynamic average of the optical conductivity at $x = 0.8$ (c) and $x = 0.2$ (d) for different temperatures.}
\label{fig:fully_terminated_MC_canonical}
\end{figure}

In Figure.~\ref{fig:fully_terminated_MC_canonical}(a), we show the configurational heat capacity as a function of temperature for \ch{Ti3C2O}$_{2-x}$\ch{F}$_x$, with $x$ ranges from 0.2 to 1.0. The peak in the configurational heat capacity indicates the order-disorder phase transition. In the ordered phase (below the phase transition temperature), the ensemble is dominated by a small number of low-energy configurations, whereas in the disordered phase (above the phase transition temperature), each termination site can be randomly occupied by \ch{-O}, \ch{-F} groups, with the probability determined by their relative chemical composition. As $x$ increases (more \ch{-F} groups), the transition temperature also increases, suggesting that higher \ch{-F} content make it thermodynamically less favorable for the surface terminations to enter the fully disordered phase.

Besides, the temperature dependence of electrical conductivity further reveals the impact of surface termination disorder, as shown in Figure~\ref{fig:fully_terminated_MC_canonical}(b). For lower \ch{-F} amount ($x = 0.2, 0.4, 0.6$, purple, green, and blue lines respectively), the electrical conductivity has a negative temperature dependence over the temperature range 100-1000~K. In contrast, for higher amount ($x = 0.8, 1.0$, red and black lines respectively), the electrical conductivity exhibits a peak, with positive temperature dependence at lower temperatures and negative dependence at higher temperatures. Additionally, the conductivity peak occurs close to the order-disorder phase transition temperature, suggesting different charge transport mechanism could dominant in the ordered and disordered phases.

In contrast, the optical conductivity exhibits a different trend. In Figure~\ref{fig:fully_terminated_MC_canonical}(c, d), we show the optical conductivities for $x = 0.8, 0.2$ at different temperatures. For both stoichiometries, the optical conductivity remains stable across the entire temperature range, even above the transition temperature. This robustness implies that optical conductivity is insensitive to the local disorder in surface terminations. Notably, we observe a prominent peak around 1.5~eV in the optical conductivity, whose magnitude decreases as $x$ increases. This implies that the optical conductivity is primarily determined by the overall chemical composition of the surface terminations, rather than the local configurations related to disorder.

Furthermore, we extended the MC simulations to the grand canonical ensemble, where the total number of surface termination groups remains fixed and the relative ratio of \ch{-O} and \ch{-F} groups are allowed to vary during the simulations. We calculated the chemical potential of \ch{O} atoms from the free energy of gas-phase \ch{O2} under standard conditions, and chose four different chemical potential values corresponding to the pressure ratios of $p_{\text{F}_2} / p_{\text{O}_2} = \{10^{-5}, 0.1, 1, 10\}$. Among these, the condition of $p_{\text{F}_2} / p_{\text{O}_2} = 10^{-5}$ represents an environment with extremely low \ch{F2} gas concentration, similar to ambient atmosphere conditions.

\begin{figure}[htb]
\centering
\includegraphics[width=\linewidth]{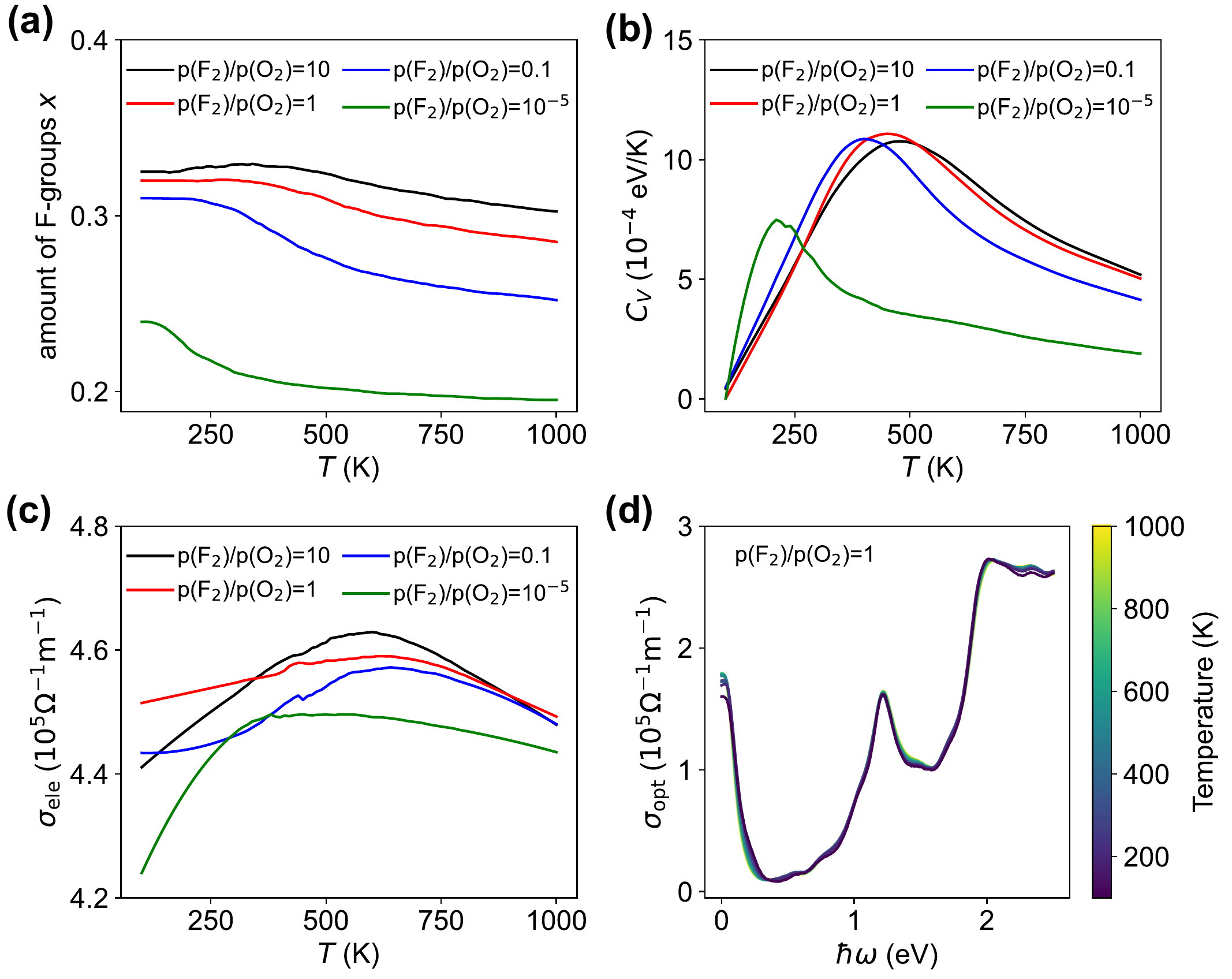}
\caption{The thermodynamic property and the ensemble averaged electrical and optical conductivities of the fully terminated MXene \ch{Ti3C2O}$_{2-x}$\ch{F}$_{x}$ under the grand canonical ensemble. (a) The average \ch{-F} content $x$ for chemical potentials corresponding to different conditions $p_{\text{F}_2} / p_{\text{O}_2} = \{10^{-5}, 0.1, 1, 10\}$. (b, c) The calculated heat capacity (b) and average electrical conductivity for different chemical potentials conditions. (d) The average optical conductivity at different temperature for the condition $p_{\text{F}_2} / p_{\text{O}_2} = 1$.}
\label{fig:fully_terminated_MC_grand_canonical}
\end{figure}

In Figure.~\ref{fig:fully_terminated_MC_grand_canonical}(a), we show the evolution of the amount of \ch{-F} groups with different temperature. Throughout the whole temperature range, \ch{-O} is thermodynamically more favorable than the \ch{-F} group, with a small compositional change ($\Delta x \leq 0.05$). While such a small composition change might suggest that results under the grand canonical ensemble would be similar to those under the canonical ensemble with fixed composition, we observe significant differences in both the thermodynamic properties and the ensemble-averaged electrical conductivity. As shown in Figure.~\ref{fig:fully_terminated_MC_grand_canonical}(b), the order-to-disorder phase transition (peak of the heat capacity spectrum) for $p_{\text{F}_2} / p_{\text{O}_2} = \{0.1, 1, 10\}$ occur at higher temperatures than those for canonical ensemble simulations with comparable \ch{-F} fraction ($x = 0.2$ or $x = 0.4$). Similarly, the electrical conductivity under the grand canonical ensemble (Figure~\ref{fig:fully_terminated_MC_grand_canonical}(c)) exhibits a peak for all chemical potential conditions, while such behavior is absent for $x = 0.2, 0.4$ under the canonical ensemble (Figure~\ref{fig:fully_terminated_MC_canonical}(b)). In contrast, the optical conductivity (Figure~\ref{fig:fully_terminated_MC_grand_canonical}(d)) is largely unaffected by the surface termination disorder and exhibits similar trend to that under the canonical ensemble. These results again demonstrate that electrical conductivity is sensitive to local atomic orderings while the optical conductivity is robust against surface termination disorder. They also highlight the flexibility of our framework in handing both canonical and grand canonical ensemble simulations.

To understand the chemical origin of the trends of electrical and optical conductivity under disorder, we consider MXene monolayers with pristine (undisordered) surface terminations, namely \ch{Ti3C2O2} and \ch{Ti3C2F2} (see Supplementary Information~S1 for details). For pristine MXene monolayers \ch{Ti3C2T2}, their structural and electronic properties have been widely studied by DFT methods~\cite{Fang24MXeneTransport, Hu17MXeneSurfaceTerminationOrigin, Dahlqvist24MXeneSurfaceTermination}. Particularly, the electronic states near the Fermi level for \ch{Ti3C2T2} MXene are predominantly contributed by the $3d$ electrons of Ti. As a result, different surface terminations do not significantly affect the band dispersions close to the Fermi level, but modulate the electron filling and shift the Fermi energy. Specifically, since \ch{F} atom contain one more electron than \ch{O} atom, the Fermi energy of \ch{Ti3C2F2} is higher than \ch{Ti3C2O2}, making \ch{Ti3C2F2} metallic and \ch{Ti3C2O2} semimetallic (with electron and hole pockets at the Fermi energy), as shown in Supplementary Figure 1. Besides, the electronic properties of other termination groups (e.g., \ch{-OH}, \ch{-Cl}, \ch{-H}) are similar to those of the \ch{-F} termination due to similar bonding configuration with the \ch{Ti} atoms, justifying our choice of using \ch{-F} and \ch{-O} as representative surface termination groups to understand the effect of disorder. 

The contrast in electronic character between \ch{Ti3C2F2} and \ch{Ti3C2O2} leads to distinct charge transport behavior. The electrical conductivity of \ch{Ti3C2O2} shows a positive temperature dependence, arising from its semimetallic nature, while \ch{Ti3C2F2} has an opposite trend, consistent with conventional metals, as shown in Supplementary Figure 1. Therefore, when the monolayer with mixed surface termination \ch{Ti3C2O}$_{2-x}$\ch{F}$_x$ is in the ordered phase, where only few configurations dominate in probability in the configurational space, the isolated \ch{-F} groups effectively act as scattering centers that disrupt the periodic potential (similar to point defects) and affect the charge transport. Therefore, the temperature-dependence of electrical conductivity is still similar to that of pristine \ch{Ti3C2O2} (positive dependence), but the magnitude of electrical conductivity is lower compared with pristine \ch{Ti3C2O2}. As temperature increases, the system undergoes phase transition into the disordered phase, where each surface termination site can be occupied by both \ch{-O} and \ch{-F} groups with probability determined by the chemical composition. In this case, the system recovers the translational symmetry on average, and \ch{-F} groups effectively raise the Fermi level and enhance metallicity of the system. This leads to a negative temperature-dependence of the electrical conductivity, consistent with charge transport in metals.

On the other hand, the different electron filling between \ch{Ti3C2F2} and \ch{Ti3C2O2} also results in distinct optical conductivity characteristic, as shown in Supplementary Figure 1. Specifically, the optical conductivity of \ch{Ti3C2O2} exhibits a peak at 1.5~eV. By analyzing the joint density of states and the magnitude of velocity matrix elements in Kubo-Greenwood formula~\cite{Xu20CoSi}, this peak originates from the interband transition between bands near the $\Gamma$ point~\cite{Fang24MXeneTransport}. However, this transition is Pauli blocked in \ch{Ti3C2F2} due to different electron occupation, resulting in the absence of peak at 1.5~eV. Unlike electrical conductivity that is mostly contributed by intraband transitions, the interband transition is insensitive to local atomic disorder that affects carrier scattering and introduces localized states, explaining the robustness of optical conductivity against disorder. Besides, with more \ch{-F} groups, the Fermi level increases such that this interband transition is optically forbidden, resulting in the absence of the peak at 1.5~eV. Therefore, unlike electrical conductivity, optical conductivity serves as a characteristic spectral feature of chemical composition rather than local atomic disorder for surface terminations of MXenes. 

\paragraph{Partially Terminated MXene \ch{Ti3C2T}$_x$}
To further explore the role of surface termination disorder, we introduce the surface termination vacancies into the configurational space. Building upon the previous fully terminated MXene \ch{Ti3C2O}$_{2-x}$\ch{F}$_{x}$ dataset, we generated an additional 2,000 \ch{Ti3C2O}$_{2-x-y}$\ch{F}$_{x}$ configurations, where the amount of \ch{-F} group ($x$) and vacancies ($y$) are randomly sampled within the ranges of $0 < x < 1$ and $0 < y < 0.3$, respectively. This results in a final dataset comprising 3,000 configurations. 

Although GNNs have been applied to predict various material properties such as total energy and band gap, they are ineffective in automatically identifying vacancies in materials and capturing their information due to the over-smoothing issue of GNN and the dilution of defect-aware features. To enhance the model's ability to capture defect-related information, we incorporate persistent homology features into the node features. As demonstrated in our previous work~\cite{Fang25Persistent}, these topological descriptors can effectively encode the local chemical environment information for each node, and more importantly the information of vacancies, including the number and positions of each vacancy. Therefore, persistent homology features can help identify vacancies and significantly improve the prediction of defect-sensitive properties, such as defect formation energies, without substantially increasing the computational cost.

Furthermore, since the vacancy positions are known \textit{a priori} during the MC simulation, we explicitly represent termination vacancy sites as virtual nodes (dummy elements) in the graph representation. This approach further enhances the model's capability to learn the underlying relationship between surface termination vacancies and target properties (energy and conductivity spectra). However, we note that such a strategy is not applicable when predicting defect-related properties in an arbitrary or unseen defective structure, where vacancy positions are not predetermined. 

In Supplementary Information S4, we show the benchmark results on the effect of virtual node representation and persistent homology features on model performance. When virtual nodes are not used, the inclusion of persistent homology features significantly improves the prediction accuracy of electrical conductivity, while the improvement for energy and optical conductivity predictions is minor. By incorporating information about the local chemical environment into the node features, the model learns better the underlying relation between electrical conductivity and structures and gains more knowledge about the electrical conductivity. This result implies that electrical conductivity is highly sensitive to variations in local atomic configurations. On the dataset of partially terminated MXene \ch{Ti3C2O}$_{2-x-y}$\ch{F}$_{x}$, our final model, which integrates both persistent homology features and virtual node representations, achieves an $R^2$ of 0.99, 0.87, 0.98 and a MAPE of 0.02\%, 3.48\%, 6.88\% for energy, optical conductivity, and electrical conductivity, respectively. Detailed comparison of the energy, electrical and optical conductivities between the DFT and GNN methods can be found Supplementary Information~S5. The strong agreement validates the downstream application to MC simulations to compute the thermodynamic and ensemble properties.

\begin{figure}[htb]
\centering
\includegraphics[width=\linewidth]{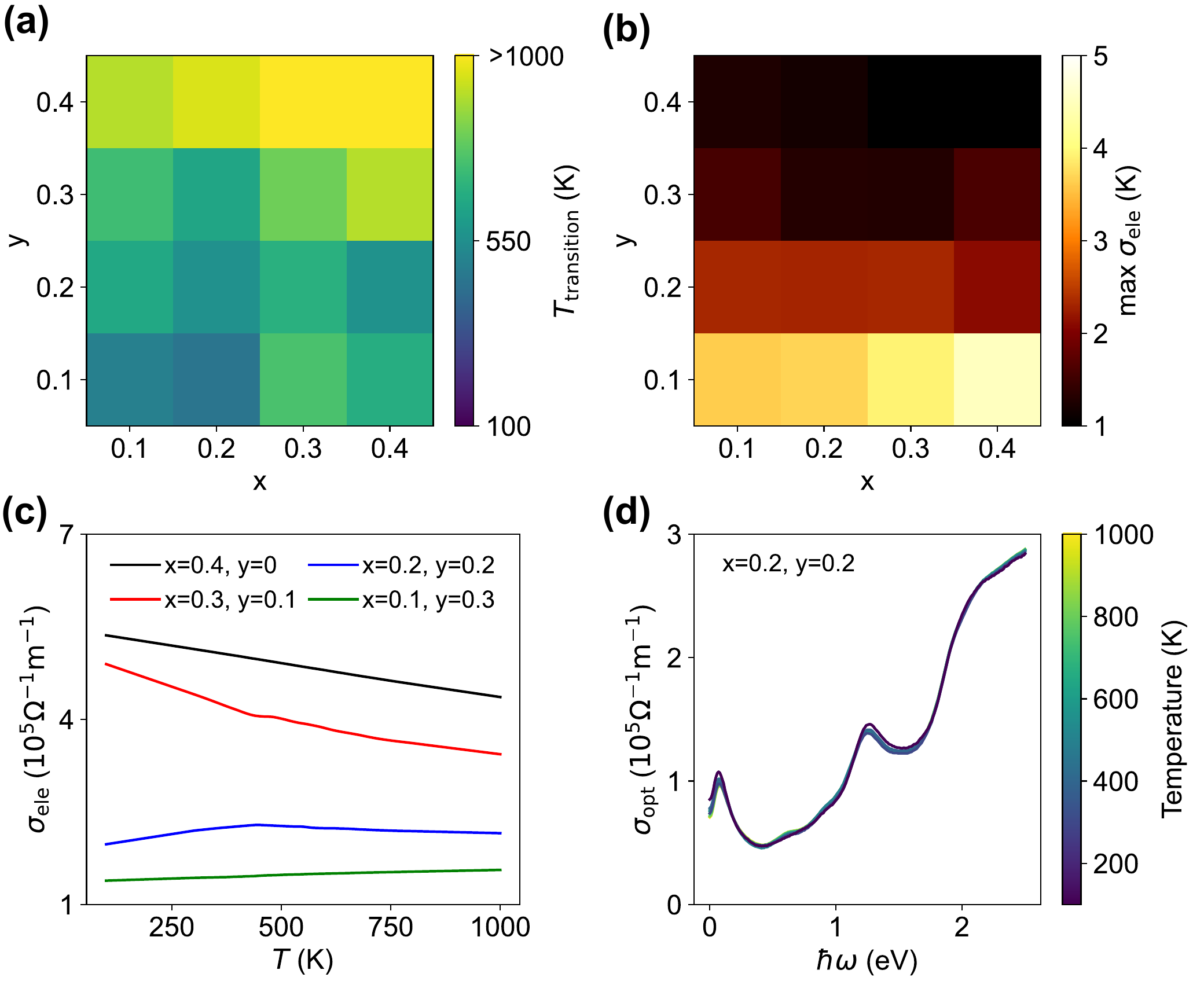}
\caption{The thermodynamic property and the ensemble averaged electrical and optical conductivities of the partially terminated MXene \ch{Ti3C2O}$_{2-x-y}$\ch{F}$_{x}$ under the canonical ensemble. (a) The calculated order-disorder phase transition temperature for different amount of\ch{-F} group ($x$) and vacancy ($y$). (b) The calculated maximum electrical conductivity for different $x$ and $y$. (c) Examples of the electrical conductivity spectra for the chemical composition with $x + y = 0.4$. (d) The average optical conductivity at different temperature for $x = y = 0.2$.}
\label{fig:partially_terminated_MC_canonical}
\end{figure}

With the trained GNN model, we performed MC simulations within the canonical ensemble on the MXene monolayer \ch{Ti3C2O}$_{2-x-y}$\ch{F}$_{x}$ for various combinations of $x$ and $y$. In Figure.~\ref{fig:partially_terminated_MC_canonical}(a), we show the order-disorder phase transition temperature, at which the configurational heat capacity attains its maximum. Overall, the phase transition temperature increases with larger values $x$ and $y$. This trend suggests that the inclusion of surface termination vacancies increases the energy barrier to enter the fully disordered phase, thus stabilizing partially ordered phases over a wide range of temperatures.

We also examined the effect of surface termination vacancy on electrical conductivity. In Figure~\ref{fig:partially_terminated_MC_canonical}(b), we show the maximum electrical conductivity with different $x, y$, and in Figure~\ref{fig:partially_terminated_MC_canonical}(c), we show representative electrical conductivity spectra for configurations with $x + y = 0.4$ (so that the fraction of \ch{-O} groups is fixed at $2 - x - y = 1.6$). These results demonstrate the reduction of electrical conductivity with more vacancies, indicating that termination vacancies can hinder the in-plane charge transport in MXenes. This arises from the disruption of the periodic potential of crystals, which modifies the band structures near the Fermi level and affects charge transport.

Furthermore, the optical conductivity for the system with $x = y = 0.2$ is shown in Figure.~\ref{fig:partially_terminated_MC_canonical}(d). Similar to the case of fully terminated MXenes, optical conductivity is robust against the presence of disorder and remains unaffected as temperature increases. In Supplementary Information~S6, we show the conductivity spectra for other $x$ and $y$ values, calculated at 300~K. With more \ch{-F} groups, the peak at 1.5~eV gradually reduces, consistent with its origin in interband transitions involving electronic states near the Fermi level, as discussed above. Additionally, with more surface termination vacancies, the optical conductivity spectra gets broadened. This behavior can be effectively viewed as the scattering effect, similar to the increasing broadening parameter used in optical conductivity calculations.

Finally, we performed the MC simulations under the grand canonical ensemble. Due to the high binding energy of \ch{-O} and \ch{-F} surface terminations, however, the amount of surface termination vacancy remains zero at all temperatures, as can be seen in Supplementary Figure 4. Therefore, under the gas-phase annealing conditions, surface termination vacancy formation is thermodynamically unfavorable. Nevertheless, under other annealing conditions, such as aqueous or electrochemical environments, we expect that different chemical potentials for \ch{-O} and \ch{-F} could lead to nonzero equilibrium vacancy amount. These results reveal the critical role of surface termination vacancies in affecting the thermodynamic and transport properties of disordered MXenes, and demonstrate the utility of our GNN-based framework in capturing such complex phenomenon.

\section*{Conclusions}
In conclusion, we present a data-driven computational framework that combines graph neural networks (GNNs) and Monte Carlo (MC) simulations to efficiently capture disorder effects and compute the ensemble-averaged functional properties of materials. By representing crystal structures as graphs, GNNs enable rapid and accurate predictions of various configurational properties, including scalar targets (e.g., formation energy) and spectral properties (density of states, conductivity spectra, etc.) across large configurational spaces. The GNN is integrated into the MC simulation to evaluate these properties for each proposed configuration during the simulation process. Thermodynamic properties and ensemble-averaged properties are then calculated using standard statistical mechanics approaches.

We apply this framework to the disordered MXene monolayer \ch{Ti3C2O}$_{2-x-y}$\ch{F}$_{x}$, showcasing its capability to capture the effect of surface termination disorder. Our results reveal that electrical conductivity is highly sensitive to local chemical environment and the surface termination disorder, with the \ch{-F} termination groups acting as scattering centers and dopants that influence the in-plane charge transport across the order-disorder phase transition. Surface termination vacancies further enhance the charge scattering, resulting in a significant reduction in the electrical conductivity. In contrast, optical conductivity is robust against the surface termination disorder and remains largely unaffected by the local structural variations, and thus reflects the global chemical compositions of the surface terminations. These results demonstrate the capability of integrating machine learning and statistical methods to quantify the thermodynamics and transport behaviors of disordered materials, offering a new route towards the theoretical modeling and rational design of complex disordered systems.

\section*{Computational Methods}
\subsection*{First-Principles Calculations}
The first-principles calculations were performed using the Vienna Ab initio Simulation Package (VASP)~\cite{Kresse93VASP1, Kresse96VASP2}. We used the projector augmented wave pseudopotentials and the Perdew-Burke-Ernzerhof functional in the generalized gradient approximation for all calculations~\cite{Kresse99PAW1, Blochl94PAW2, Perdew96GGA}. The kinetic energy cutoff for the plane-wave basis sets is 550~eV, and the energy convergence threshold is $10^{-7}$~eV. For static calculations, we used the $\mathbf{k}$-point grid with a density of 0.03 $2 \pi$/\AA), and a vacuum level of 20~\AA. To describe the electron correlation effect, we employed the DFT+U method~\cite{Anisimov91DFT_U}, with a $U$ value of 3 eV is added to the $3d$ orbitals of Ti atoms. We also included the DFT-D3 correction to take account of the van der Waals interactions~\cite{Grimme11DFTD3}.

\subsection*{High-Throughput Wannierization Workflow}
To calculate the electrical and optical conductivities of configurations, we employed the high-throughput Wannierzation workflow based on maximally localized Wannier functions, as developed in our previous work~\cite{Fang25Database}. We analyzed the projected density of states near the Fermi level and chose the atomic orbitals with leading contributions as the projection functions to construct the Wannier functions. We further performed an energy scan to search for optimal energy windows that reproduce the electronic states near the Fermi level.

After obtaining the Wannier Hamiltonian, we calculated the optical conductivity $\sigma_\text{opt}^{\alpha\beta} (\hbar \omega)$ according to the Kubo-Greenwood formula~\cite{Animalu67OpticalConductivity} and electrical conductivity $\sigma_\text{ele}^{\alpha\beta} (T)$ according to the Boltzmann transport theory~\cite{Pizzi14TransportProperties}
\begin{align}
    \sigma_\text{opt}^{\alpha\beta} (\hbar \omega) &= \frac{i e^2 \hbar}{(2 \pi)^3} \sum_{nm} \int  \frac{f_{m\mathbf{k}} - f_{n\mathbf{k}}}{E_{m\mathbf{k}} - E_{n\mathbf{k}}} \frac{v_{nm}^\alpha (\mathbf{k}) v_{mn}^\beta (\mathbf{k})}{E_{m\mathbf{k}} - E_{n\mathbf{k}} - (\hbar \omega + i \eta)} d\mathbf{k} \label{eqn: optical_conductivity} \\
    \sigma_\text{ele}^{\alpha\beta} (T) &= e^2 \int_{-\infty}^{\infty} \left( -\frac{\partial f(E, T)}{\partial E} \right)\Sigma^{\alpha\beta} (E) dE \label{eqn:electrical_conductivity}
\end{align}
where
\begin{align}
    \Sigma^{\alpha\beta} (E) = \frac{1}{V} \sum_{n, \mathbf{k}} v^\alpha (n, \mathbf{k}) v^\beta (n, \mathbf{k}) \delta(E - E_{n\mathbf{k}}) \tau
\end{align}
is the transport distribution function tensor. Here, $\alpha, \beta = \{ x, y, z \}$ represents the Cartesian directions, $v_{nm} (\mathbf{k})$ is the velocity matrix element, $f_{m\mathbf{k}}$ and $E_{m\mathbf{k}}$ represent the Fermi-Dirac occupation and the band energies, and $\tau = 10$~fs was used as the constant relaxation time when calculating the electrical conductivity~\cite{Zhou20RelaxationTime, Sernelius91RelaxationTime}. To accelerate convergence, we used the broadening factor of 0.05~eV for the Delta function. Since MXenes are two-dimensional materials, the optical and electrical conductivities in this work all refer to the average of the $xx$- and $yy$-components.

\subsection*{GNN Model and Training}
Our GNN model features a typical global feature regression architecture. The input crystal structure of each configuration is converted into a graph, where atoms and chemical bonds are represented as nodes and edges. The node features are chosen as the one-hot encodings for each chemical species, and the edge features are chosen as the bond length expanded in the Gaussian basis. We used the equivariant GNN framework, which outputs the global feature of the graph, whose length is equal to the sum of the lengths of energy (since energy is a scalar, its length is 1), optical conductivity, electrical conductivity. The global feature is then sliced into these three targets directly. Note that we used only one GNN model to predict the three targets simultaneously, instead of using three different models.

With the energy, optical conductivity, and electrical conductivity obtained from DFT and GNN methods, we calculated the mean absolute error (MAE, see Supplementary Information S3 for details) for those targets. The three MAE values are summed together into the total loss function for the model during training. We also tried multi-task optimization techniques, such as dynamical weight averaging and GradNorm, but found no significant improvement on our task.

To train the GNN model, we splited the dataset into the training, validation, and testing set according to 60:20:20 ratio. The mean absolute error (MAE) on the validation set was tracked during the training stage, and the model with the minimum MAE on the validation set was used for testing. We optimized the hyperparameters based on the Bayesian optimization method, implemented in Optuna~\cite{Optuna, Bergstra11TPE}. The set of hyperparameters used in this work includes the number of equivariant convolution layers, the number of embedding channels, the multiplicity of the convolution layers, learning rate, weight decay, and batch size.
 
\subsection*{MC Simulation}
With the trained GNN model, we performed the MC simulation with the Metropolis sampling method. We sampled the temperature from 1,000~K to 100~K, with the interval of 10~K. At each individual temperature, we performed the simulation with $2 \times 10^5$ steps; thus the whole simulation contains $1.82 \times 10^7$ MC steps. For each temperature, we discard the first 20\% steps and only use the later 80\% steps to compute the configurational heat capacity, as well as the thermodynamic average of electrical and optical conductivities. The configurational heat capacity is calculated 
\begin{align}
    C_v = \frac{\langle E^2 \rangle - \langle E \rangle^2}{k_\text{B} T^2} \label{eqn:heat_capacity}
\end{align}
where $\langle E \rangle$ and $\langle E^2 \rangle$ are the average energy and average squared energy. The average conductivity is calculated as the harmonic mean of the conductivity for each configuration,
\begin{align}
    \sigma_\text{ave} = \frac{N}{\sum_i \frac{1}{\sigma_i}} \label{eqn:conductivity_average}
\end{align}
where $\sigma_i$ is the conductivity for each configuration $i$, and $N$ is the total number of configurations that are averaged.



\section*{Supporting information}
Supplementary Information S1 to S6\\
Supplementary Tables 1 to 3\\
Supplementary Figures 1 to 4\\

\section*{Acknowledgments}
Z.F. thanks helpful advice from Dr. Tian Qiu and Dr. Zhen Jiang for helpful discussions on MC simulations. This work is supported by the U.S. Department of Energy, Office of Science, Basic Energy Sciences, under Award No. DE-SC0023664. This research used resources of the National Energy Research Scientific Computing Center (NERSC), a U.S. Department of Energy Office of Science User Facility located at Lawrence Berkeley National Laboratory, operated under Contract No. DE-AC02-05CH11231 using NERSC award BES-ERCAP0029544.

\section*{Competing interests}
The authors declare no competing interests.

\bibliography{bibliography}


\end{document}